\documentclass[manuscript]{aastex}
\newcommand{\myemail}{jerome.petri@astro.unistra.fr}

\shorttitle{Mass and moment of inertia of neutron stars}
\shortauthors{J. P\'etri}

\begin{document}
  
\title{Constraining the mass and moment of inertia of neutron stars
  from quasi-periodic oscillations in X-ray binaries.}

\author{J\'er\^ome P\'etri}
\affil{Observatoire Astronomique de Strasbourg, 11 rue de
  l'Universit\'e, 67000 Strasbourg, France.}

\email{\myemail}

\begin{abstract}
  Neutron stars are the densest objects known in the Universe. Being
  the final product of stellar evolution, their internal composition
  and structure is rather poorly constrained by measurements.

  It is the purpose of this paper to put some constrains on the mass
  and moment of inertia of neutron stars based on the interpretation
  of kHz quasi-periodic oscillations observed in low mass X-ray
  binaries.

  We use observations of high-frequency quasi-periodic observations
  (HF-QPOs) in low mass X-ray binaries (LMXBs) to look for the average
  mass and moment of inertia of neutron stars. This is done by
  applying our parametric resonance model to discriminate between slow
  and fast rotators.

  We fit our model to data from ten LMXBs for which HF-QPOs have been
  seen and the spin of the enclosed accreting neutron star is known.
  For a simplified analysis we assume that all neutron stars possess
  the same properties (same mass $M_*$ and same moment of inertia
  $I_*$).  We find an average mass $M_* \approx 2.0-2.2\, M_{\odot}$.
  The corresponding average moment of inertia is then $I_* \approx 1-3
  \times 10^{38}\;{\rm kg\,m^2} \approx 0.5-1.5 \, (10\;\textrm{
    km})^2 \, M_\odot$ which equals to dimensionless spin parameter
  $\tilde{a} \approx 0.05-0.15$ for slow rotators (neutron stars with
  a spin frequency roughly about 300~Hz) respectively $\tilde{a}
  \approx 0.1-0.3$ for fast rotators (neutron stars with the spin
  frequency roughly about 600~Hz).
\end{abstract}


\section{Introduction}

Neutron stars are excellent astrophysical laboratories to test matter
above nuclear density \citep{2006ARNPS..56..327P}.  Unfortunately,
there is nowadays no way for nuclear physicists to investigate matter
at such extremely high densities in laboratories.  Moreover, because
of the lack of knowledge about the behavior of particles in these
extreme regimes, there is yet no consensus on a satisfactory equation
of state for nucleons. Many modern equations of state have been
proposed, based on non-relativistic approximations or with help on
relativistic field theory (see \cite{2007ASSL..326.....H} and
references therein). These equations of state at or above nuclear
density predict different mass to radius relations for neutron stars.
The answer or a piece of it could maybe come not from terrestrial
laboratories but from the sky \citep{2007PhR...442..109L,
  2010arXiv1002.3153O}.  Indeed, it has been claimed that measuring
the mass and the radius of neutron stars will help to constrain the
proposed equations of state and to reject some of them
\citep{1998ApJ...508..791M, 2007Ap&SS.308..371L}.

HF-QPOs observations in LMXBs is a unique tool to test gravity in the
strong field regime and to learn about the behavior of particles at
high densities.  Further detailed observations and modelling of QPOs
for individual objects will help getting more insight into the
properties of individual accreting neutron stars.

How can we then estimate their mass and radius? In binary neutron
stars showing up as pulsars, the task is relatively easy
\citep{1999ApJ...512..288T}. The very accurate clock furnished by the
pulsar serves as an efficient instrument to deduce the orbital motion
and other parameters in this system \citep{2006AdSpR..38.2721N}. Such
techniques have been successfully applied by numerous authors, finding
masses aggregating around $1.4\,M_\odot$ (for a summary, see e.g.
\cite{2007PhR...442..109L}).

For neutron stars in LMXBs, the situation is less favorable although
some attempts have been made.  For a recent review on different
methods to compute neutron star parameters, see for instance
\cite{2010arXiv1001.1642B} or also \cite{2007MNRAS.374..232Z}.  During
their life, neutron stars in binaries accrete matter from their
companion, an amount which can reach a substantial fraction of their
initial mass. Not surprisingly, their final mass can deviate
significantly from the fiducial~$1.4\,M_\odot$.  Actually, high-mass
neutron stars seem plausible with $M\approx1.6-1.9\,M_\odot$
\citep{2006MNRAS.373.1235C}.  For some pulsars like SAX~J1808.4-3658,
such high masses were also found \citep{2008MNRAS.391.1619D}.

Quasi-periodic oscillations (QPOs) seen in LMXBs can help to diagnose
motion in strong gravitational fields and maybe solve the problem of
determination of mass and radii.  Although their estimates and related
quantities are strongly model-dependent, picking up a particular
QPO-model certainly helps on making some strong assertions about
neutron star properties, see~\cite{1998ApJ...508..791M} and
\cite{2009AAZ}.

For a thorough review on X-ray variability and QPOs, see
\cite{2006csxs.book...39V}.  Several models for high-frequency and
low-frequency QPOs have been proposed as described in this review.
Some invoke resonance mechanisms~\citep{2007RMxAC..27...18K}, other
relativistic precession motion \citep{1999PhRvL..82...17S} or MHD
Alfven waves~\citep{2004A&A...423..401Z, 2005A&A...436..999R}.
However, some important problems are still unsolved
\citep{2007RMxAC..27....8A}.

In the present work, we show how to compute the average mass and
moment of inertia of neutron stars by fitting kHz-QPO observations in
LMXBs for slow and fast rotators.  This paper is divided in two main
parts. In Sec.\ref{sec:Model} we briefly summarize the parametric
resonance model, more details can be found in
\cite{2005A&A...439L..27P} and in \cite{2005A&A...443..777P}. In
Sec.\ref{sec:Results}, we apply the model to ten LMXBs and deduce
their key parameters. The conclusions are presented in
Sec.\ref{sec:Conclusion}. Finally, Appendix~\ref{sec:Appendix}
discusses the way to extend the model to allow for variable QPO
frequencies, an important point with respect to observations.

\section{Model and method}
\label{sec:Model}

In this section, we recall the main results of the model. The
essential feature is the presence of a rotating neutron star which
does not possess an axial symmetry about its rotation axis. The origin
of the asymmetry can be due to the magnetic field tilted with respect
to the rotation axis or due to anisotropic and inhomogeneous stellar
interior, producing either a rotating asymmetric magnetic or
gravitational field. \cite{2005A&A...439L..27P} has shown that this
induces some driven motion in the accretion disk due to a parametric
resonance. Therefore the disk will show strong response to this drive
by oscillating across the equatorial plane at some given preferred
radii where the resonance condition is satisfied.  More explicitly,
remember that vertical resonance occurs whenever the vertical
epicyclic frequency is equal to the perturbation frequency as measured
in the locally corotating frame
\begin{equation}
  \label{eq:Resonance}
  m \, |\Omega(r,a_*) - \Omega_* | = 2 \, \frac{\kappa_{\rm z}(r,a_*)}{n}
\end{equation}
where $m$ is the azimuthal number of the perturbation mode,
$\Omega(r,a_*)$ the orbital frequency in the disk at radius~$r$,
$\Omega_*$ the spin of the neutron star, $a_*$ a length related to the
angular momentum $J_* = I_* \, \Omega_*$ by $a_* = J_* / M_*\,c$,
$\kappa_{\rm z}(r,a_*)$ the vertical epicyclic frequency, $I_*$ the
stellar moment of inertia and $n$ an integer.  The frequencies
$\Omega$ and $\kappa_{\rm z}$ are expressed for a test particle in
Kerr space-time.  They depend explicitly on the radius~$r$ and on the
angular momentum~$a_*$ as
\begin{eqnarray}
  \label{eq:FreqOrbitale}
\Omega(r,a_*) & = & \frac{\sqrt{G\,M_*}}{r^{3/2} + a_*\,\sqrt{R_{\rm g}}} = 
\frac{c^3}{G\,M_*} \, \frac{1}{\tilde{r}^{3/2} + \tilde{a}} \\
  \label{eq:FreqVerticale}
\kappa_{\rm z}(r,a_*) & = & \Omega(r,a_*) \,
  \sqrt{1 - 4 \, \frac{\tilde{a}}{\tilde{r}^{3/2}} + 3 \,
    \frac{\tilde{a}^2}{\tilde{r}^2}}
\end{eqnarray}
$R_{\rm g} = G\,M_*/c^2$ is the gravitational radius of the star,
$\tilde{r} = r/R_{\rm g}$ and $\tilde{a} = a_* / R_{\rm g}$.  We
explicitly used the Kerr metric to find expressions for these orbital
and epicyclic frequencies. However, this is not the best approximation
for the exterior of realistic rotating neutron stars, since the
quadrupole moment of the star usually causes large deviations from the
gravitational field that would be created by a simpler Kerr black hole
geometry. Nevertheless, the Kerr geometry was recently used by
\cite{2010ApJ...714..748T} to estimate the mass and the spin of
neutron stars from the relativistic precession model.  They argued
that the metric well describes the exterior of rotating high-mass
neutron stars and can be used when the non-rotating mass implied by
the model is in the upper interval of masses allowed by the equations
of state.  This strongly supports our approach since already for
$\tilde{a}=0$ the parametric resonance model implies a neutron star
mass which is relatively high, around $2\,M_\odot$.

Eq.~(\ref{eq:FreqVerticale}) giving the vertical epicyclic frequency
in the Kerr approximation was first published by
\cite{1981GReGr..13..899A}. Some useful properties are summarized in,
e.g., \cite{1998bhad.conf.....K} and \cite{2005A&A...437..775T}.

From the known spin of the neutron star, we can deduce its angular
moment by $J_* = I_* \, \Omega_*$, assuming a given value for the
moment of inertia~$I_*$.  Therefore, guessing a mass and a moment of
inertia, we can solve quantitatively Eq.~(\ref{eq:Resonance}) for the
orbital frequency~$\Omega$ and try to match observations of kHz-QPOs.

For slowly rotating stars, $\tilde{a}\ll1$, we retrieve the Newtonian
expression
\begin{equation}
  \Omega(r,a_*) \approx \Omega(r,0) = \kappa_{\rm z}(r,0) = \sqrt{\frac{G\,M_*}{r^{3/2}}}
\end{equation}
from which the solution of Eq.~(\ref{eq:Resonance}) follows immediately
\begin{equation}
  \Omega(r,0) = \frac{m\,n}{m\,n\pm2} \, \Omega_*
\end{equation}
The orbital frequency~$\Omega(r,0)$ should remain smaller than this at
the innermost stable circular orbit (ISCO) given in the non-rotating
limit by
\begin{equation}
  \nu_{\rm isco} = 2198\textrm{ Hz} \, \left( \frac{M_\odot}{M_*}
  \right) = 1570\textrm{ Hz} \, \left(
    \frac{M_*}{1.4\,M_\odot} \right)^{-1}
\end{equation}
This would give a first guess for the expected QPO frequencies,
knowing the mass~$M_*$. Actually, because the spin frequency is well
known from X-ray bursts for instance, we can do better and include the
angular momentum~$\tilde{a}$ into the description, but then the moment
of inertia comes in as another free parameter.

Several LMXBs have been observed with known spin rate and showing the
twin peak QPO phenomenon. Depending on the neutron star rotation
speed, they have been classified as slow rotator for
$\nu_*\lesssim400$~Hz or as fast rotator for $\nu_*\gtrsim400$~Hz,
($\nu_* = \Omega_*/2\,\pi$). For slow rotators, the twin kHz-QPO
difference, $\Delta\nu^{\rm obs} \approx \nu_*$, is almost equal to
the spin frequency while for fast rotators, it is equal to half of it,
$\Delta\nu^{\rm obs} \approx \nu_*/2$.  It is sometimes argued that
this slow/fast rotator dichotomy is an artefact.
\cite{2007MNRAS.381..790M} reexamined the data from all these sources
and claimed that there is no clear trend in any segregation between
them. They showed that the kHz QPO frequency difference $\Delta\nu$ is
much more concentrated (mostly in the window [200,400]~Hz, precisely a
Gaussian with mean 308~Hz and standard deviation 36~Hz) than the range
of neutron star spin (from 100~Hz to more than 600~Hz). Using a
Kolmogorov-Smirnov statistical test, they found it highly improbable
that $\Delta\nu$ and $\nu_*$ are correlated.  $\Delta\nu$ is almost
constant and only weakly $\nu_*$-dependent with a fit done by
\cite{2007A&A...471..381Y} who find
\begin{equation}
  \label{eq:FitQPO}
  <\Delta\nu> \approx - 0.19 \, \nu_* + 389\textrm{ Hz} .
\end{equation}
This makes the link between spin frequency and QPO frequency
difference questionable. \cite{2007MNRAS.381..790M} went even further
and made the strongest assumption of independence between both
frequencies. Such hypothesis could rule out simple resonance models
(those invoking linear oscillations and no inward motion of the flow
for instance) as they claimed. However, this conclusion is not exactly
true and would not hold anymore if some simplifying assumptions of any
resonance model are left. Non-linear effects as well as a radially
inward motion of the accretion disk can significantly change the
oscillation frequency which becomes a function of the amplitude of
oscillations and explicitly on time because the proper orbital, radial
and vertical epicyclic frequencies vary when matter approaches the
neutron star surface.  This was discarded so far. However, in this new
extended picture, the spin frequency still plays an important role by
triggering the resonance at some preferred radius, bringing the disk
into off-plane oscillations that are slowly advected by the flow and
drift downwards to the neutron star. Therefore, $\nu_*$ does not give
a clear imprint to the precise kHz-QPO frequencies as it seems (not)
seen in the data, but serves to launch the mechanism.  Moreover, these
motions occur due to matter flow influenced by gravity in a strong
field regime, and thus the ISCO plays a central role. It is not the
purpose of this paper to study the drifting and non-linear terms,
which will deserve full attention in another work.  Here, to give a
taste, we only draw the basic lines of the consequences of these
effects in Appendix~\ref{sec:Appendix}.

Any model predicting a fixed frequency ratio faces difficulties to
explain the data since this frequency ratio is not only 3/2 or 4/3
(where most of the observations cluster), but covers a wider range as
seen by \cite{2005A&A...437..209B}. Although a strong linear
correlation exists, it differs significantly from the 3/2 ratio, see
\cite{2005A&A...437..209B} and \cite{2005AN....326..864A,
  2005ragt.meet....1A}. In addition, the frequency ratio clustering
around the 3/2 value first found by \cite{2003A&A...404L..21A} could
be well explained by a uniform distribution of the lower and upper kHz
QPO set in the source.  The 3/2 peak in the observed ratio
distribution comes from selection effects (sensitivity of measurement
tools) since there is only a very narrow range of frequency ratio
where both QPOs are sufficiently strong in order to be detected. The
details can be found in works of
\cite{2008AcA....58...15T,2008AcA....58..113T} and
\cite{2010MNRAS.401.1290B} who elaborated this issue.  We emphasize
that their results do not contradict the parametric resonance model.
The question of the viability of such models remains fully open and
subject to strong debates. Moreover, \cite{2008NewAR..51..835B} looked
carefully at 4U1820-30 and found a gap of roughly 100~Hz in the QPO
frequency distribution that is not attributed to selection effect and
sharply peaked around a 4/3~ratio. For this special binary, it seems
that some frequencies are disfavored. In other words, within the
orbital QPO interpretation, preferred radii exist within the disk,
supporting the resonance model.

When the star rotates slowly, its geometrized angular
momentum~$\tilde{a}$ remains small and a first order expansion
for~$\nu_{\rm isco}(\tilde{a})$ with respect to~$\tilde{a}$ is
possible. In the next section, we will show how to use this linear
approximation to find severe constrains on the stellar mass and moment
of inertia.

Another way to tackle the resonance condition
Eq.~(\ref{eq:Resonance}), in the general case for
arbitrary~$\tilde{a}$, is to work directly with the full expressions
given by Eq.~(\ref{eq:FreqOrbitale})-(\ref{eq:FreqVerticale}). This
requires a numerical algorithm to search for the allowed frequencies
and is also done in the next section, Sec.~\ref{sec:Results}.

A dozen LMXBs have been inventoried to exhibit the above mentioned
behavior. The LMXBs sample used to fit our model for the kHz-QPO
difference as measured by some other authors are summarized in
Table~\ref{tab:Data} with appropriate references.
\begin{table}
  \centering
  \begin{tabular}{cccc}
    \hline
    &   separation~$\Delta\nu$ (in Hz) & spin~$\nu_*$ (in Hz) & ratio~$\Delta\nu/\nu_*$\\
    \hline
    \hline
    Millisecond pulsars &  & &  \\
    XTE J1807-294       & 179-247 & 191  & 0.94-1.29 \\
    SAX J1808.4-3658    &   195   & 401  & 0.49 \\
    \hline
    Atoll sources & & & \\
    4U 1608-52          & 224-327 & 619  & 0.36-0.53 \\
    4U 1636-53          & 217-329 & 581  & 0.37-0.57 \\
    4U 1702-43          &   333   & 330  & 1.01 \\  
    4U 1728-34          & 271-359 & 363  & 0.75-0.99  \\ 
    KS 1731-260         &    266  & 524  & 0.51   \\
    4U 1915-05          & 290-353 & 270  & 1.07-1.31  \\ 
    IGR J17191          &   330   & 294  & 1.12   \\
    SAX J1750.8-29      &   317   & 601  & 0.53   \\
    \hline
  \end{tabular}
  \caption{The detail of the LMXBs with known twin kHz QPOs
    and spin frequencies. For the source of the data, see for instance
    \cite{2005A&A...437..209B, 2007MNRAS.tmp..101B,
      2006MNRAS.366.1373Z, 2007A&A...471..381Y, 2006AdSpR..38.2675V}
    and references therein.}
  \label{tab:Data}
\end{table}
We want our model to adjust to this set as close as possible by
looking for appropriate mass and moment of inertia. Let us take an
index~$i$ tracing this set of LMXBs by writing $i\in(LMXBs)$.  For
each binary, the observed twin kHz-QPO frequency difference is known
as~$\Delta\nu_i^{\rm obs}$. Fixing~$M_*$ and $I_*$, we get a predicted
$\Delta\nu_i^{\rm model}$ from our parametric resonance model.  To
evaluate the goodness of our fit, we introduce a merit function
$\mathcal{F}$ defined by summation over all the LMXBs and compare the
discrepancy between predicted and measured QPO differences,
such that
\begin{equation}
  \label{eq:MeritFunction}
  \mathcal{F} = \sum_{i\in (LMXBs)} \left| \frac{\Delta\nu_i^{\rm obs} - 
      \Delta\nu_i^{\rm model}}{\sigma_i} \right|
\end{equation}
with a statistical weight~$\sigma_i$.  The summation should be
understood over the set of observed systems.  $\Delta\nu_i^{\rm
  obs/model}$ are the observed/predicted HF-QPO frequency difference
and $\sigma_i$ the error in the observed QPO frequency difference for
the binary labeled~$i$. We use the $L_1$-norm but other choices are
possible like the $L_2$-norm, although the latter being less robust.
We found no significant changes when applying the second choice. In
any case, the best fit corresponds to a minimum of the figure-of-merit
function~$\mathcal{F}$. Moreover, we tried other merit functions with
no difference in the best fit parameters.

Finally, some words about the mass-dependence on stellar rotation.
Assuming the same mass as well as the same moment of inertia for all
the set of neutron star binaries is a crude first guess. A detailed
description of the inner structure of rapidly rotating neutron stars
is a difficult calculation only numerically treatable
\citep{1983bhwd.book.....S, 2007coaw.book.....C}. For uniform
rotation, the mass increase is expected to be less than 20\%. However,
\cite{2003ApJ...583..410L} computed equilibrium configurations with
differential rotation and found an increase up to 60\%, even for
moderate spin rate. The salient feature to keep in mind from all these
studies is an increase of the gravitational mass with rotation.

Thus, to better adjust the observations without handling all these
complicated computations, we can however release the constant mass
hypothese and use a neutron star spin dependent mass based on the
following heuristic argument \citep{2007rapp.book......}. The rotation
of the neutron star, containing a fixed number of~$N$ nucleons,
increases its gravitational mass~$M(N,\Omega_*\neq0)$ compared to the
non rotating limit~$M(N,\Omega_*=0)$. Because kinetic energy is
equivalent to mass and therefore induces gravitation, a simple
relation between both gravitational masses is such that
\begin{equation}
  \label{eq:RelationSpinMasse}
  M(N,\Omega_*) \, c^2 = M(N,0) \, c^2 + \frac{1}{2} \, I_* \, \Omega_*^2
\end{equation}
For the remainder of the paper, we use lighter notations, setting $M
\equiv M(N,0)$ for the mass of a non-rotating neutron star and $M_*
\equiv M(N,\Omega_*)$ for that of the same neutron star (i.e. equal
number of baryons~$N$) but rotating at an angular speed~$\Omega_*$.
Eq.~(\ref{eq:RelationSpinMasse}) shows the quadratic dependence on
spin~$\Omega_*^2$, the same functional dependence as the one from the
study of \cite{1968ApJ...153..807H}. The relative mass correction is
therefore
\begin{equation}
  \label{eq:CorrectionMasse}
  \frac{\delta M}{M_\odot} = \frac{I_* \, \Omega_*^2}{2\,c^2\,M_\odot}
  = 1.76 \times 10^{-3} \, \left(
    \frac{I_*}{10^{38} \textrm{ kg\,m}^2} \right) \, \left(
    \frac{\nu_*}{400 \textrm{ Hz} } \right)^2
\end{equation}
So it remains small even for fast rotators.

\section{Results}
\label{sec:Results}

The above model and fitting technique is now applied to the dozen of
fast and slow rotators. We emphasize that in order to make any
prediction on the mass and moment of inertia, we have to take the same
properties for the whole sample of accreting neutron stars in the
observed LMXBs. Indeed, adopting different parameters for each system
could significantly change the orbital and epicyclic frequencies. Most
importantly, the frequency at the ISCO, scaling like $1/M_*$ would
change from one binary to another. But in our segregation between slow
and fast rotators, the precise value of the orbital frequency at the
ISCO is the salient feature to interpret the abrupt change in the twin
peak frequency difference. A varying $M_*$ would shift this sharp
transition to lower or higher frequencies from one binary to another.
Thus the zero-th order choice, to highlight the general trend, is to
keep the same mass for all neutron stars.

Let us first give an estimate for the gravitational mass~$M_*$ and
moment of inertia~$I_*$ along the following arguments.  The
geometrized spin parameter~$\tilde{a}$ is defined as
\begin{equation}
  \label{eq:MomentCinetique}
  \tilde{a} = \frac{I_*\,\Omega_*\,c}{G\,M_*^2} = 0.145 \, \left(
    \frac{I_*}{10^{38} \textrm{ kg\,m}^2} \right) \, \left(
    \frac{\nu_*}{400 \textrm{ Hz} } \right) \, \left( \frac{M_*}{1.4\,M_\odot} \right)^{-2}
\end{equation}
From this expression it is clear that it remains small compared to
unity, and this even for fast rotators.  In this case, to first order
in~$\tilde{a}$, the orbital frequency at the ISCO is
\citep{1985ApJ...297..548K, 1990ApJ...358..538K, 1998ApJ...508..791M}
\begin{equation}
  \nu_{\rm isco}(\tilde{a}) = 2198 \textrm{ Hz} \, ( 1 + 0.75 \, \tilde{a} ) \,
  \frac{M_\odot}{M_*}
\end{equation}
Assuming that $\tilde{a}\le0.3$, an inaccuracy introduced by this
simplification with respect to the Kerr solution (due to neglecting
higher order terms in $j$) is smaller than $10\%$.  We put explicitly
the spin dependence through~$\tilde{a}$ on the left hand side for
latter convenience.  And therefore the relation between $M_*$ and
$I_*$ becomes approximately
\begin{equation}
\label{eq:IM}
I_* = \frac{2}{3\,\pi} \, \frac{G\,M_*^2}{c\,\nu_*} \, \left[ \left(
    \frac{\nu_{\rm isco}(\tilde{a})}{2198 \textrm{ Hz}} \right) \, \frac{M_*}{M_\odot} - 1 \right]
\end{equation}
According to our parametric resonance model, for slowly rotating
stars, the twin kHz-QPOs are given by $\nu_1^{\rm s} = 2\,\nu_*$
and $\nu_2^{\rm s} = 3\,\nu_*$, where the superscript~s stands for
{\it slow}. Because $\nu_1^{\rm s},\nu_2^{\rm s}$ are interpreted as
the frequencies of the orbital motion, they need to be less than that
at the ISCO
\begin{equation}
  \label{eq:SlowRotator}
  \nu_1^{\rm s},\nu_2^{\rm s} \le \nu_{\rm isco}
\end{equation}
For increasing spin of the neutron star $\nu_*$, at some point,
$\nu_2^{\rm s}$ will approach and eventually overtake~$\nu_{\rm
  isco}$.  Thus $\nu_2^{\rm s}$ will be forbidden as a HF-QPO.  As a
consequence, the next two dominant twin kHz-QPOs are identified as
$\nu_1^{\rm f} = 1.5\,\nu_*$ and $\nu_2^{\rm f} = 2\,\nu_*$.
Therefore, the QPO frequency difference $\Delta\nu / \nu_* =
(\nu_2-\nu_1) / \nu_*$ jumps suddenly from 1.0 and 0.5.  According to
the data taken from \cite{2007MNRAS.381..790M, 2008AIPC.1068..163V},
this should happen in the neutron star spin range
$\nu_*\in[363,401]$~Hz. This is probably the most salient feature in
the slow against fast rotator discrepancies.  Fitting these data
requires that the switching from slow to fast rotator occurs for
neutron star spin between 363~Hz and 401~Hz. More precisely, for
$\nu_*\le363$~Hz, $\Delta\nu / \nu_* \approx1$ which we interpret as
no effect on motion in the observable disk from the presence of an
ISCO. This implies that $\nu_{\rm isco}(363\textrm{ Hz}) \ge 3 \,
\nu_*=1089$~Hz, we put the spin rate into coma to distinguish between
different rotators, an essential remark for our constrains.  Next, for
$\nu_* \ge 401$~Hz, $\Delta\nu / \nu_* \approx 0.5$ which we interpret
as a clear signature of the ISCO. This implies that $\nu_{\rm
  isco}(401\textrm{ Hz}) \le 3\,\nu*=1203$~Hz. Express in terms of the
ISCO, the transition from slow to fast rotator should happen when the
two conditions below are satisfied
\begin{eqnarray}
  \label{eq:nuISCOrange1}
  \nu_{\rm isco}(363\textrm{ Hz}) & \ge & 1089 \textrm{ Hz} \\
  \label{eq:nuISCOrange2}
  \nu_{\rm isco}(401\textrm{ Hz}) & \le & 1203 \textrm{ Hz}.
\end{eqnarray}
This condition supplemented with the relation Eq.~(\ref{eq:IM}) sets
two constrains on $M_*$ and $I_*$, an allowed region in the
$(M_*,I_*)$ plane. Next, a third bound for the couple $(M_*,I_*)$ is
possible along the following lines. For fast rotators, the ISCO is
clearly taken into account. But for the highest accreting system
with~$\nu_*=619$~Hz, the ratio is still $\Delta\nu / \nu_* \approx
0.5$, the upper kHz-QPO being $\nu_2^{\rm f} = 2\,\nu_*=1238$~Hz and
the lower kHz-QPO being $\nu_1^{\rm f} = 1.5\,\nu_* = 929$~Hz. We
conclude that {\it for this particular system}
\begin{equation}
   \label{eq:nuISCOdernier} 
   \nu_{\rm isco}(619\textrm{ Hz}) \ge 1238 \textrm{ Hz}.
\end{equation}
The last and general constrain is that there is no naked singularity
in the Kerr metric or stated mathematically, $|\tilde{a}|\le1$. In
terms of the moment of inertia, it means that
\begin{equation}
   \label{eq:singularite} 
  I_* \le \frac{G\,M_*^2}{2\,\pi\,c\,\nu_*}.
\end{equation}
The less favorable case (most restrictive one) corresponds to
$\nu_*=619$~Hz. This leads to
\begin{equation}
   \label{eq:singularite2} 
   \left( \frac{I_*}{10^{38}\textrm{ kg m}^2} \right) 
   \le 2.26 \, \left( \frac{M_*}{M_\odot} \right)^2.
\end{equation}
For later use, we introduce $I_0 = 10^{38}\textrm{ kg m}^2$.
\begin{figure}
  \centering
  \includegraphics[width=0.45\textwidth]{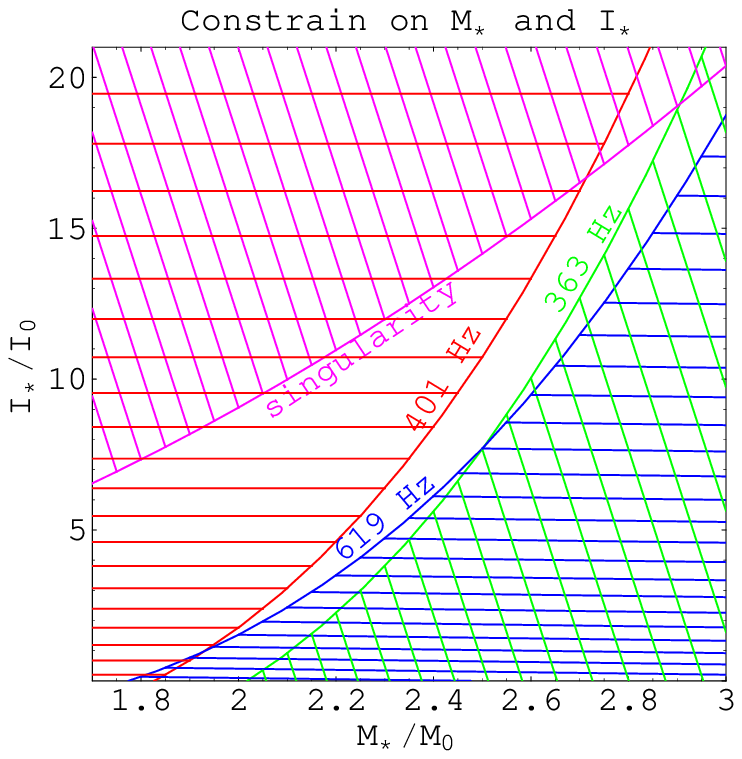}
  \caption{The four constrains
    Eq.~(\ref{eq:nuISCOrange1})-(\ref{eq:singularite}), in the
    $(M_*,I_*)$-plane, labeled with the spin frequency or marked by
    {\it singularity} for Eq.~(\ref{eq:singularite}). The hashed
    regions are forbidden. The minimum allowed mass and moment of
    inertia are around $M \approx 1.93\,M_\odot$ and $I_* \approx
    0.98\,I_0$.}
  \label{fig:ContrainteMI}
\end{figure}
All these constrains,
Eq.~(\ref{eq:nuISCOrange1})-(\ref{eq:singularite}), are summarized and
shown in a $(M_*,I_*)$-plane depicted in Fig.~\ref{fig:ContrainteMI}.
The hashed regions are forbidden and only a small area in white
survives around the first diagonal in the figure. This plot clearly
emphasizes the existence of a lower and upper bound for both the mass
and moment of inertia. We found the minimum values to be $ M_{*\rm
  min} = 1.9\,M_\odot$ and $I_{*\rm min} = 0.98\,I_0$ whereas the
maximum ones are $M_{*\rm max} = 2.9\,M_\odot$ and $I_{*\rm max} =
19.1\,I_0$.  Neutron star structure models predict $I_*$ close to or
slightly above $I_0$ so that we will favor the lower bounds and expect
masses in the vicinity of $1.9\,M_\odot$.

In a last step, we use the figure-of-merit function $\mathcal{F}$,
Eq.~(\ref{eq:MeritFunction}), to fit the data and also the variation
of mass with spin rate according to Eq.~(\ref{eq:RelationSpinMasse}).
We span a vast range in the $(M,I_*)$-plane to compute the merit
function. Note however the subtil change in unknowns compared to the
previous analytical study. Now we use a constant mass for the
{\it non-rotating limit}, $M$, instead of a constant mass for the rotating
star, $M_* = M + I_*\,\Omega_*^2/2\,c^2$.  Actually, the discrepancy
between both approaches is small and can be neglected at the end of
the study.

\begin{figure}
  \centering
  \includegraphics[width=0.45\textwidth]{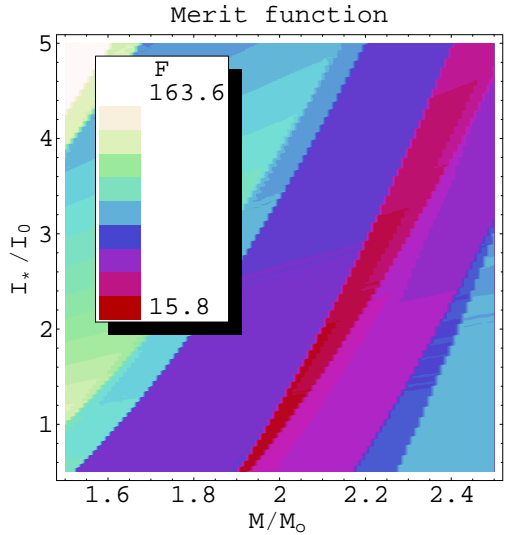}
  \caption{Isocontours of the merit function~$\mathcal{F}$ vs the fiducial mass
    of the non-rotating neutron star~$M$, normalized to the solar
    mass~$M_\odot$, and vs its moment of inertia~$I_*$, normalized to
    $I_0 = 10^{38}\; {\rm kg\,m^2}$. The minimum value of~$f$ lies
    around $M \approx 2.0-2.2\,M_\odot$ and $I_* \approx 1-3\,I_0$.}
  \label{fig:MeritFunction1}
\end{figure}

\begin{figure}
  \centering
  \includegraphics[width=0.45\textwidth]{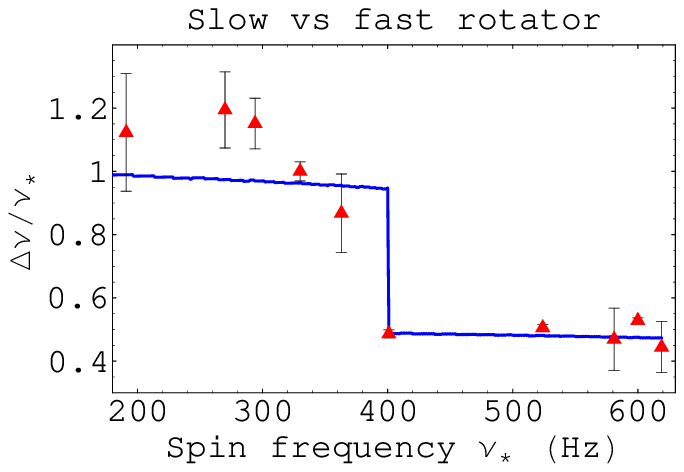}
  \caption{Observations of slow and fast rotators (red triangles) and
    fit obtained by our model (blue solid line). The best fit
    parameters are $M \approx 2.0-2.2 \, M_{\odot}$ and $I_* \approx 1-3\,I_0$.}
  \label{fig:Rotateur1}
\end{figure}

The results presenting the isocontours of the merit function are
summarized in Fig.~\ref{fig:MeritFunction1}. The resulting region for
minimization of~$\mathcal{F}$ is marked in red and shapes very
similarly to the previous diagram $(M_*,I_*)$,
Fig.~\ref{fig:ContrainteMI}. The most probable mass and moment of
inertia are $M \approx 2.0-2.2\,M_\odot$ and $I_* \approx 0.5-1.5 \,
(10\;\textrm{ km})^2 \, M_\odot$.  The best fit according to these
values is shown in Fig.~\ref{fig:Rotateur1} where the spin rate is
plotted on the x-axis and the twin kHz-QPOs difference normalized to
the spin rate is plotted on the y-axis.  First, we retrieve the
segregation between slow and fast rotators at the correct frequency as
expected. Next, for fast spinning stars, the theoretical curve agrees
very well with observations. Nevertheless, for slow rotation rates,
the spread around unity is significant and cannot be explained in a
straightforward way by our predictions.  Clearly, some refinement of
the model is still needed and under investigations, including other
aspects of the plasma flow around an accreting magnetized neutron
star.

\section{Discussion and Conclusion}
\label{sec:Conclusion}

In this paper, we investigated further the consequences of forced
oscillations induced in an accretion disk to explain the twin kHz-QPOs
in LMXBs. Our model is able to discriminate between slow and fast
rotator as already shown in \cite{2005A&A...439L..27P}. Moreover, with
help on new data from a dozen rotators, we were able to constrain the
average mass and moment of inertia of neutron stars. We found for the
best fit $M \approx 2.0-2.2 \, M_\odot$ and $I_* \approx 0.5-1.5 \,
(10\;\textrm{ km})^2 \, M_\odot$. Whereas the moment of inertia gives
roughly the same value as those obtained from independent ways by
solving the stellar structure with several equations of state
\citep{2008ApJ...685..390W}, the neutron star mass appears rather
large.  This effect could be an artefact of its constancy from one
binary system to another.  Better fits suggests to look at each system
individually and remove the constant mass approximation for the whole
set of LMXBs, leading to a spread in the mass distribution function
for neutron stars. But this requires a much more detailed separate
analysis of each binary with their own specificities (accretion rate,
magnetic field strength for instance) and better observations.  New
time analyzing instruments like the HTRS (High Time Resolution
Spectrometer) project on board IXO will give more insights into
supra-nuclear matter and strong gravity physics
\citep{2008SPIE.7011E..10B}.


\appendix
\section{Toward a more realistic model}
\label{sec:Appendix}

So far our {\it linear} parametric resonance model predicts fixed
radii where the resonance conditions Eq.~(\ref{eq:Resonance}) are
satisfied. Therefore the orbital frequencies remain also constant,
leading to fixed kHz-QPOs.  This can only be an approximation since
oscillations are non-linear and gas or particles drift slowly towards
the neutron star due to accretion.  In other words, advection
increases the orbital and vertical epicyclic frequencies and puts the
system (particles) out of resonance.  Actually, if non-linear terms
are retained, the proper frequency of the vertical excursions depends
on the amplitude of these oscillations.  Therefore, it is possible
that the excitation and proper frequencies {\it adjust themself} in
such a way to maintain the oscillator in high amplitude motion. We
call this {\it a parametric auto-resonance mechanism}. Its explanation
is given in more details along the following lines.

The idea of non-linear resonance to explain QPO observations has
already been discussed by many authors, see for instance
\cite{2004PASJ...56..553R, 2004ragt.meet...91H, 2005AN....326..824H,
  2003PASJ...55..467A}.  Although they considered a resonance between
oscillatory modes different from those relevant to the model presented
within this paper, their results show also how non-linear phenomena
can drastically improve their models.

Let us see how parametric auto-resonance works. Add a non-linear cubic
term in the usual Mathieu equation governing the vertical displacement
$z(t)$. Note that, to first order, a quadratic term would not lead to
a change in proper frequency with amplitude so a cubic term is more
relevant for our discussion.  Thus, our non-linear oscillator takes
the form below
\begin{equation}
  \label{eq:OscillateurNL1}
  \frac{d^2 z}{dt^2} + \kappa_{\rm z}^2(t) \, [ 1 + h \, \cos( m (
  \Omega(t) - \Omega_*) \, t ) ] \, z = \beta \, z^3
\end{equation}
$h$ is the strength of the excitation. Now, an important new fact is
that the vertical epicyclic~$\kappa_{\rm z}(t)$ and
orbital~$\Omega(t)$ frequencies are {\it time dependent}. These
variable coefficients mimic the particle radial drift.  How does the
position of the resonant particles evolve with time due to loss of
angular momentum expected from accretion? A simple argument to get the
temporal dependence is the following. Assume that the thin accretion
disk possesses a power-law axisymmetric surface density $\Sigma(r)$
such that
\begin{equation}
  \label{eq:DensiteSurface}
  \Sigma(r) = \Sigma_0 \, \left( \frac{r}{r_0} \right)^\alpha
\end{equation}
where~$\alpha$ is the power law index and $\Sigma_0$ the surface
density at~$r_0$.  The flow starts to accrete at an initial
speed~$v_0$ (directed radially inwards such that $\vec v = - v_0 \,
\vec e_r$) at radius~$r_0$. The conservation of mass implies
\begin{equation}
  \label{eq:ConsevationMasse}
  2\,\pi\,r\,\Sigma(r)\,v(r) = 2\,\pi\,r_0\,\Sigma_0\,v_0
\end{equation}
Integration with respect to time, using the fact that for a test
particle $v(r) = - dr/dt$ (projection along $-\vec e_r$), we get for
$\alpha\neq-2$
\begin{equation}
  \label{eq:Accretion1}
  r(t) = \left[ r_0^{\alpha+2} - (\alpha+2) \, r_0^{\alpha+1} \, v_0 (t-t_0) \right]^{\frac{1}{\alpha+2}}
\end{equation}
where $r_0 = r(t_0)$ is the initial position of the particle.
Specializing to a uniform density disk model, i.e. $\alpha=0$, the
particle falls onto the neutron star along the trajectory
\begin{equation}
  \label{eq:Accretion2}
  r(t) = \sqrt{ r_0^{2} - 2 \, r_0 \, v_0 \, ( t - t_0 ) }
\end{equation}
For convenience, in the remainder of this paper, we will use this
expression for the particle radial path. Furthermore, we make a shift
of time by the replacement $(t-t_0) \to t$.  This temporal dependence
on radius~$r(t)$ governs the time evolution of the orbital~$\Omega(t)$
as well as the vertical epicyclic~$\kappa_{\rm z}(t)$ frequencies.
Parametric auto-resonance will settle in if the excitation is
efficient enough to maintain phase locking.  In some special cases, we
can look for semi-analytical solutions by the method of averaging
\citep{2001PhRvE..64c6619K}.  The underlying idea is to smooth the
fastest time scale~$\Omega(t)$ and to keep track of only the secular
amplitude change and phase evolution of the oscillation, compared to
an harmonic oscillator.

\begin{figure}
  \centering
  \includegraphics[width=0.45\textwidth]{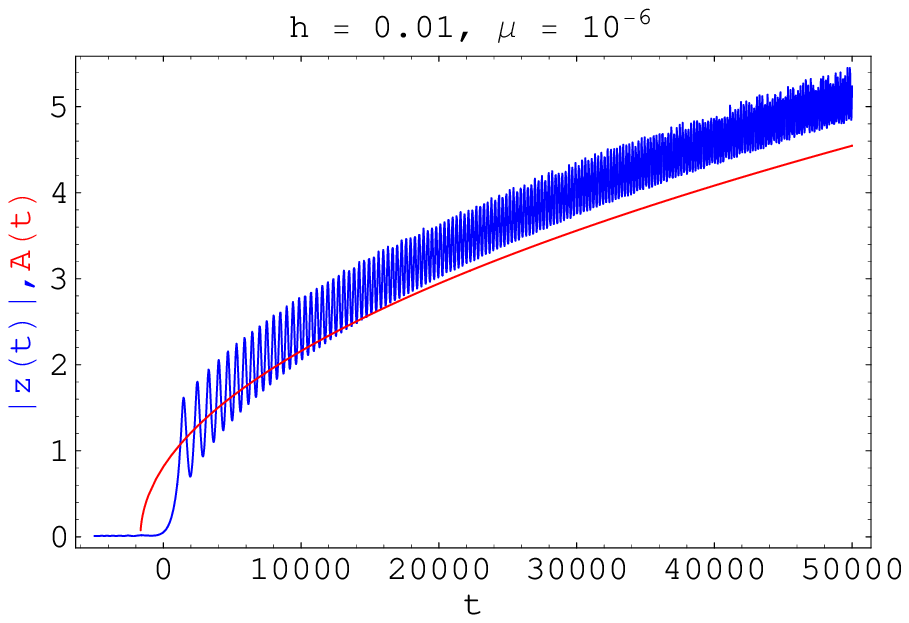}
  \caption{An example of growing oscillation amplitude $|z(t)|$, solid
    blue line, due to the parametric auto-resonance. The parameters
    are $h=10^{-2}$, $\beta=10^{-2}$ and $\mu=v_0/r_0=10^{-6}$. The
    orbital frequency is normalized to unity $\Omega=1$ or
    equivalently $\Omega_*=3$. The short orbital period of $2\,\pi$ is
    not plotted. The solid red line shows the quasi-stationary fixed
    point solution for the amplitude $A(t)$, Eq.~(\ref{eq:PointFixe}),
    in good agreement with the true solution.}
  \label{fig:Autoresonance}
\end{figure}

We apply this technique to our resonance model. First, for small
enough times $v_0\,t\ll r_0$, an expansion of radius and frequencies
yields (assuming a non-rotating body, $\tilde{a}=0$)
\begin{eqnarray}
  \label{eq:r2t}
  r(t) & \approx & r_0 - v_0 \, t \\
  \label{eq:Omega2t}
  \Omega(t) & \approx & \Omega_0 \, \left( 1 +
    \frac{3}{2} \, \frac{v_0\,t}{r_0} \right) \\
  \Omega_0 & \approx & \sqrt{\frac{G\,M}{r_0^3}} \\
  \kappa_{\rm z}(t) & \approx & \Omega(t)
\end{eqnarray}
The non-linear parametric resonance model for vertical motion becomes
\begin{equation}
  \label{eq:OscillateurNL2}
  \frac{d^2 z}{dt^2} + \Omega_0^2 \left( 1 +
    \frac{3}{2} \frac{v_0\,t}{r_0} \right)^2 \left[ 1 + h 
    \cos\left\{ m \left( \Omega_0 \, \left( 1 + \frac{3}{2}
          \frac{v_0\,t}{r_0} \right) - \Omega_* \right) t \right\} \right] z = \beta z^3
\end{equation}
Next, the method of averaging looks for solutions expanded into $z(t)
= A(t) \, \cos \vartheta(t)$, the amplitude being $A(t)$ and the phase
being $\vartheta(t)$, $A(t)$ varying on a timescale much longer than
the period of oscillations. It is preferable to introduce a new phase
defined by
\begin{equation}
  \label{eq:Phase}
  \psi = \vartheta - \Omega_0 \, t
\end{equation} 
to get rid of the fastest timescale represented by~$\Omega_0$.  Let us
have a look on the behavior of the first resonance.  Specializing to
this particular case corresponding to $n=1$ and $m=1$, the resonance
condition, in the Newtonian limit, is $\Omega_0 = \Omega_*/3$.
Inserting into Eq.~(\ref{eq:OscillateurNL2}), the vertical motion
satisfies
\begin{equation}
  \label{eq:OscillateurNL3}
  \frac{d^2 z}{dt^2} + \frac{\Omega_*^2}{9} \, \left( 1 +
    \frac{3}{2} \frac{v_0\,t}{r_0} \right)^2 \left[ 1 + h \,
    \cos\left\{ \frac{2}{3} \Omega_* \, \left( 1 - \frac{3}{4}
        \frac{v_0\,t}{r_0} \right) \, t \right\} \right] z = \beta \, z^3
\end{equation}
The straightforward way is to solve numerically this second order
differential equation with appropriate initial conditions.  We show an
example of numerical integration of Eq.~(\ref{eq:OscillateurNL3}) in
Fig.~\ref{fig:Autoresonance} when the particle enter in the first
resonance, $n=m=1$. The full solution, $z(t)$, is not plotted because
of the small timescale, whereas the evolution of the amplitude is
shown by a solid blue line, $|z(t)|$. A clear increase of the
amplitude with respect to time is demonstrated.  This shows that
resonance can occur even for variable orbital and excitation
frequencies.
Averaging provides another mean to solve approximately for the
amplitude $A$ and the phase $\psi$. For accretion timescales much
longer than the orbital motion, which is usually the case, these new
unknowns evolve according to
\begin{eqnarray}
  \label{eq:Average}
  \frac{dA}{dt} & = & \frac{h \, A \, \Omega_* \, \sin(2\,\psi)}{12} \\
  \frac{d\psi}{dt} & = & \frac{h \, \Omega_* \, \cos(2\,\psi)}{12} -
  \frac{9\,\beta\,A^2}{8\,\Omega_*} + \frac{\Omega_*}{2} \, \frac{v_0\,t}{r_0}
\end{eqnarray}
A stationary solution corresponds mathematically speaking to a fixed
point satisfying $\frac{dA}{dt} = \frac{d\psi}{dt} = 0$. Inverting
this system for the phase and amplitude, we find the real positive
amplitude by
\begin{eqnarray}
  \sin(2\,\psi) & = & 0 \\
  \label{eq:PointFixe}
  A(t) & = & \frac{2}{3} \,\Omega_* \, \sqrt{ \frac{1}{\beta} \, \left(
      \frac{v_0\,t}{r_0} + \frac{h}{6} \right)  }
\end{eqnarray}
This analytical solution, solid red line in
Fig.~\ref{fig:Autoresonance}, agrees very well with the
straightforward numerical integration of
Eq.~(\ref{eq:OscillateurNL3}).  From that, we conclude that
oscillations grow slowly as time goes.  Simultaneously, the orbital
period increases according to Eq.~(\ref{eq:Omega2t}) and consequently
so does the kHz-QPO frequency. We recall that this describes the
short time evolution of the particle just after entering resonance. In
this limit, the fractional increase in the frequency remains small, a
few percent.  Other resonance, with different $m,n$ will follow the
same trend.  Accretion and non-linearities allow the QPO peaks to
drift towards higher frequencies and release the fixed frequency ratio
result.

The analysis of its long term evolution (with the full non-linear time
dependence retained) requires more investigation and is left for
future work.  This discussion aimed at emphasizing that the parametric
resonance model extended to more realistic situations met in accretion
disks (non linearity and chirped frequency) can explain an increase in
the observed kHz-QPO frequencies by {\it parametric auto-resonance}.

To conclude, representative references about the real behaviour of the
QPO amplitudes and coherence times can be found in
\cite{2005MNRAS.361..855B, 2005AN....326..808B, 2006MNRAS.371.1925M}.
Finally, we refer to \cite{2009A&A...497..661T} and
\cite{2009A&A...499..535H} for a discussion on the relation between
the strength of the twin QPOs and its possible explanation within the
framework of a non-linear resonance theory.

The study done in this appendix is very preliminary and will be
included in an extended model of our parametric resonance.

\end{document}